\documentclass[lettersize,journal]{IEEEtran}
\usepackage{amsmath,amsfonts}
\usepackage{algorithmic}
\usepackage{array}
\usepackage[caption=false,font=normalsize,labelfont=sf,textfont=sf]{subfig}
\usepackage{textcomp}
\usepackage{stfloats}
\usepackage{url}
\usepackage{amssymb}
\usepackage{verbatim}
\usepackage{graphicx}
\usepackage{caption}
\usepackage{subfig}
\usepackage{xcolor}

\def\BibTeX{{\rm B\kern-.05em{\sc i\kern-.025em b}\kern-.08em
    T\kern-.1667em\lower.7ex\hbox{E}\kern-.125emX}}
\usepackage{balance}
\begin{document}
\title{Measurement of Dielectric Loss in Silicon Nitride at Centimeter and Millimeter Wavelengths
}

\author{%
Z. Pan, P. S. Barry, T. Cecil, C. Albert, A. N. Bender, C. L. Chang, R. Gualtieri, J. Hood, J. Li, J. Zhang, M. Lisovenko, V. Novosad, G. Wang, and V. Yefremenko

\thanks{
(\textit{Corresponding author: Zhaodi Pan)}}
\thanks{Z. Pan is with the Argonne National Laboratory, 9700 South Cass Avenue, Lemont, IL, 60439, USA (email: panz@anl.gov)}%
\thanks{P. Barry is with Cardiff University, Cardiff CF10 3AT, UK (email: barryp2@cardiff.ac.uk)}
\thanks{T. Cecil, A. N. Bender, R. Gualtieri, J. Li, J. Zhang, M. Lisovenko, V. Novosad, G. Wang, and V. Yefremenko are with the Argonne National Laboratory, 9700 South Cass Avenue, Lemont, IL, 60439, USA (emails: cecil@anl.gov; abender@anl.gov; rgualtieri@anl.gov; juliang.li@anl.gov; jianjie-zhang@outlook.com; mlisovenko@anl.gov; novosad@anl.gov; gwang@anl.gov; yefremenko@anl.gov)}%
\thanks{C. Albert was with the the University of Chicago, Chicago, IL 60637 USA and is now with Caltech, 1200E California Blvd, Pasedena CA 91125 USA (email: calbert@caltech.edu)}
\thanks{C. L. Chang is with the Argonne National Laboratory, Argonne, IL 60439 USA and the University of Chicago, 5640 South Ellis Avenue, Chicago, IL 60637 USA (email: clchang@kicp.uchicago.edu) }%
\thanks{J. Hood is with the University of Chicago, 5640 South Ellis Avenue, IL 60637 USA (email: hood.asro@gmail.com) }%

}

\markboth{Journal of \LaTeX\ Class Files,~Vol.~0, No.~0, April~2023}%
{How to Use the IEEEtran \LaTeX \ Templates}

\maketitle

\begin{abstract}
This work presents a suite of measurement techniques for characterizing the dielectric loss tangent across a wide frequency range from $\sim$1~GHz to 150 GHz using the same test chip. In the first method, we fit data from a microwave resonator at different temperatures to a model that captures the two-level system (TLS) response to extract and characterize both the real and imaginary components of the dielectric loss. The inverse of the internal quality factor is a second measure of the overall loss of the resonator, where TLS loss through the dielectric material is typically the dominant source. The third technique is a differential optical measurement at 150~GHz. The same antenna feeds two microstrip lines with different lengths that terminate in two microwave kinetic inductance detectors (MKIDs). The difference in the detector response is used to estimate the loss per unit length of the microstrip line. Our results suggest a larger loss for SiN$_x$ at 150~GHz of $\boldsymbol{\mathrm{\tan \delta\sim 4\times10^{-3}}}$ compared to $\boldsymbol{\mathrm{2.0\times10^{-3}}}$ and $\boldsymbol{\mathrm{\gtrsim 1\times10^{-3}}}$ measured at $\sim$1~GHz using the other two methods. {These measurement techniques can be applied to other dielectrics by adjusting the microstrip lengths to provide enough optical efficiency contrast and other mm/sub-mm frequency ranges by tuning the antenna and feedhorn accordingly. } 

\end{abstract}


\section{Introduction}
Dielectric materials are used extensively in mm-wave superconducting detectors for astrophysical or cosmological surveys. Experiments with photometer arrays such as CMB-S4 use microstrip lines to transfer the mm-wave signal received at the antenna to the detectors, where dielectric loss limits the overall optical efficiency \cite{abitbol2017cmb}. Dielectric loss is more critical for on-chip spectroscopy, where it constrains the quality factor of the on-chip filter bank, which limits the combination of spectral resolving power and the optical efficiency \cite{hailey2016low}. The lumped-element MKID architecture typically utilize an interdigitated capacitor, but adapting to parallel-plate capacitors can significantly reduce the size {of the resonators} and increase their packing density. However, one disadvantage associated with parallel-plate capacitors is that more of the power is confined to the thin-film dielectric layer; thus, the loss mechanism in dielectrics will limit detector quality factor {and increase} the noise. In quantum information, the dielectric loss can cause decoherence \cite{martinis2005decoherence}. The defining roles of low-loss dielectrics in multiple applications motivate this study. 

A few low-loss dielectrics have been developed and are commonly in use, including $\mathrm{SiO_2}$ \cite{li2013improvements}, $\mathrm{SiN_x}$ \cite{ye2019low}, amorphous Si (a-Si) \cite{defrance2022low}, and silicon carbide \cite{buijtendorp2021hydrogenated}. State-of-art loss tangent of $O(\mathrm{10^{-5}})$ has been achieved for a-Si \cite{defrance2022low} and silicon carbide \cite{buijtendorp2021hydrogenated} at microwave frequencies. However, past measurements mostly focused on a few GHz, while applications for mm-wave astrophysical surveys deal with radiation at millimeter wavelengths. Though a few measurements at mm-wave have been reported, more data {is needed} to validate the dielectric performance across different deposition techniques at mm-wave and cross-compare with measurements at a few GHz. {Given the discrepancy between cm and mm-wavelengths that a few groups observed for SiN$_x$, hydrogenated amorphous Si and SiC~\cite{hubmayr2022optical, buijtendorp2022hydrogenated, hahnle2021superconducting}},  more work is needed to study the discrepancy and elucidate other potential loss mechanisms at mm-wavelengths beyond the dielectrics.

This paper presents our recent effort to measure the dielectric loss from centimeter to millimeter wavelengths. We designed and fabricated a test structure that can probe dielectric loss using three methods: 1) fitting frequency vs. temperature response to a TLS model, 2) inverting the quality factor in the low power limit, and 3) measuring the differential loss for detectors coupled via different lengths of microstrip transmission lines. {Compared to the Fabry–Pérot method in \cite{hahnle2021superconducting}, our method doesn’t require a coherent source that can sweep in frequency and only needs an internal blackbody to feed the antenna. The detector chip has a simple design but offers three different ways to measure the dielectric loss.} In a previous paper \cite{hood2022testing}, we presented a design where the signal from a paddle of the ortho-mode transducer (OMT) probe is split into two transmission lines with different lengths and coupled to two detectors. However, the detectors did not show a noticeable optical response. We then simplified the optical coupling and fed more power to each detector by differentiating the signal from a pair of OMT paddles and removing the power split. The new strategy showed the expected optical coupling level. Here, we report on progress made having resolved several technological issues since \cite{hood2022testing}: 1) developed a more robust electrical contact between the aluminum inductor and niobium capacitor, 2) solving the step coverage when the  transmission line climbs up onto the capacitor pads, 3) fixed unwanted shorts from niobium etching residuals, and 4) improving parasitic leakage of the optical signal from neighbouring detectors. 

\section{Device Design and Fabrication}
Measurement devices optimised for mm-wave were fabricated using a similar process to that reported in \cite{hood2022testing} but with a few key differences. Devices start as a 150~mm silicon wafer with a commercially-grown 450~nm thermal SiO$_2$ and 2~$\mu$m {low-pressure chemical-vapor deposited (LPCVD)} SiN$_x$ bilayer. These layers form the released membrane for the OMT probes. A 300~nm thick Nb ground plane is deposited and patterned by fluorine-based reactive ion etch (RIE). Next, a 500~nm-thick layer of SiN$_x$ is deposited for the microstrip dielectric. The SiN$_x$ was deposited by chemical vapor deposition using SiH$_4$ and N$_2$ precursors at a substrate temperature of 275$^{\circ}$C. Following the microstrip dielectric, the microstrip wiring is fabricated from a 170~nm-thick Nb layer, along with a 30~nm NbN capping layer and patterned via liftoff. Next, a 50~nm Al layer is deposited and patterned into inductors via liftoff. Previously we had deposited the Al inductors before the Nb capacitors. However, due to issues with poor galvanic contact between the layers, we reversed the steps and added the NbN capping layer. The NbN limits oxidation of the Nb, making a galvanic contact to the Al more robust. Additionally, in previous devices, the Nb wiring layer was etched, which occasionally leads to shorting when the material is not removed from step sidewalls by the anisotropic RIE. Switching to using liftoff eliminated this issue. After the wiring layer processing, the OMT membrane is released by a silicon deep RIE from the backside of the wafer. We use the same etch step to separate individual chips from the wafer. The final step is an O$_2$ plasma used to remove a protective resist layer from the front side of the wafer. 

\begin{figure}[!ht]
\centering
\includegraphics[width=0.95\linewidth]{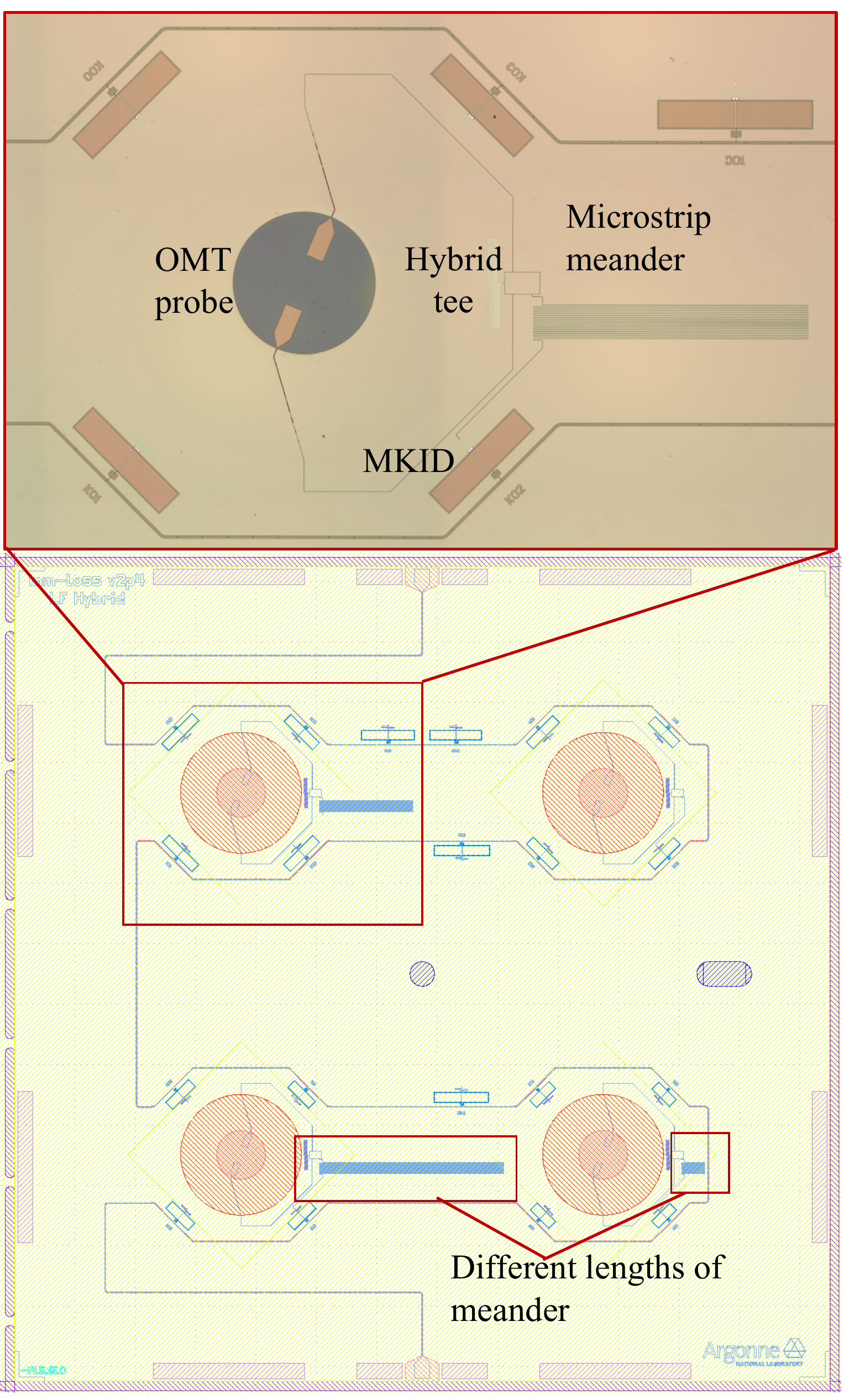}
\caption{Detector chip design (bottom) and the microscope image for one pixel (top). A CPW feedline couples to 20 MKIDs, out of which four receive optical from OMT probes through hybrid tees and different lengths of microstrip transmission lines. Other detectors do not directly connect to the antenna and are references for systematics and dark pickup. }
\label{fig:design_image}
\end{figure}

Fig.~\ref{fig:design_image} bottom and top show the chip design and a microscope image of a segment near a pixel. A coplanar waveguide (CPW) feedline is capacitively coupled to 20 MKIDs detectors, each containing a parallel-plate niobium capacitor and a meandered aluminum inductor line. We choose parallel-plate capacitors over interdigitated capacitors (IDCs) due to their higher energy filling factor $F$ that is close to unity. Four of the 20 resonators are coupled to optical signals from the OMT via hybrid tee and different lengths of microstrip meander lines. At mm-wave, the dielectric loss can be extracted by comparing the optical response of detectors connected through varying lengths of microstrip transmission lines. In \cite{hood2022testing}, we split the signal from one arm of the OMT into two detectors, which reduced the optical signal level to be comparable to systematics caused by temperature variation and noise scatter. To improve this, we increase the signal level by a factor of four by removing the signal split and differentiating signals from two OMT fins with a hybrid tee. Other detectors should be dark, meaning they do not receive signals from the antenna via microstrip. The primary purpose of the dark detectors is to monitor systematics from temperature variation and optical leakage. 

\section{Dielectric Loss from TLS Model at 1-2~GHz}
\label{sect: tls_model}
The two-level system (TLS) modeling in amorphous solids was first introduced by \cite{anderson1972anomalous} and \cite{phillips1987two}, where groups of atoms may tunnel between sites {corresponding to different energy states}. The dielectric modification by TLS can be derived assuming a log-uniform distribution of tunneling states, and the subsequent shift in resonant frequency for a resonator is given by \cite{gao2008physics}:

\begin{equation}
\label{eq:tls_model}
\frac{f(T)-f_{0}}{f_{0}}=\frac{F \delta_{\mathrm{TLS}}^0}{\pi}\left[\operatorname{Re\Psi}\left(\frac{1}{2}-\frac{\hbar f_{0}}{j k_{B} T}\right)-\log \frac{\hbar f_{0}}{k_{B} T}\right], 
\end{equation}
where $f(T)$ is the resonator frequency at temperature $T$, $f_0$ is the TLS-free resonance frequency, $F$ is the energy filling factor in the amorphous dielectrics,  $\delta_{\mathrm{TLS}}^0$ is the intrinsic TLS loss, $\Psi$ is the digamma function, and $k_B$ is the Boltzmann constant. Typically $F$ and $\delta_{\mathrm{TLS}}^0$ are degenerate when fitting $f(T)$ vs. $T$ to the model, but we eliminated the degeneracy in our system using parallel-plate capacitors with filling factors $F$ close to one. {An increase in stage temperature can break Cooper pairs and change the kinetic inductance, resulting in a downshift of the resonance frequency \cite{gao2008equivalence}. But this effect takes over after around 300~mK for aluminum and is beyond the range of our study.} 

\begin{figure}[!ht]
\centering
\includegraphics[width=\linewidth]{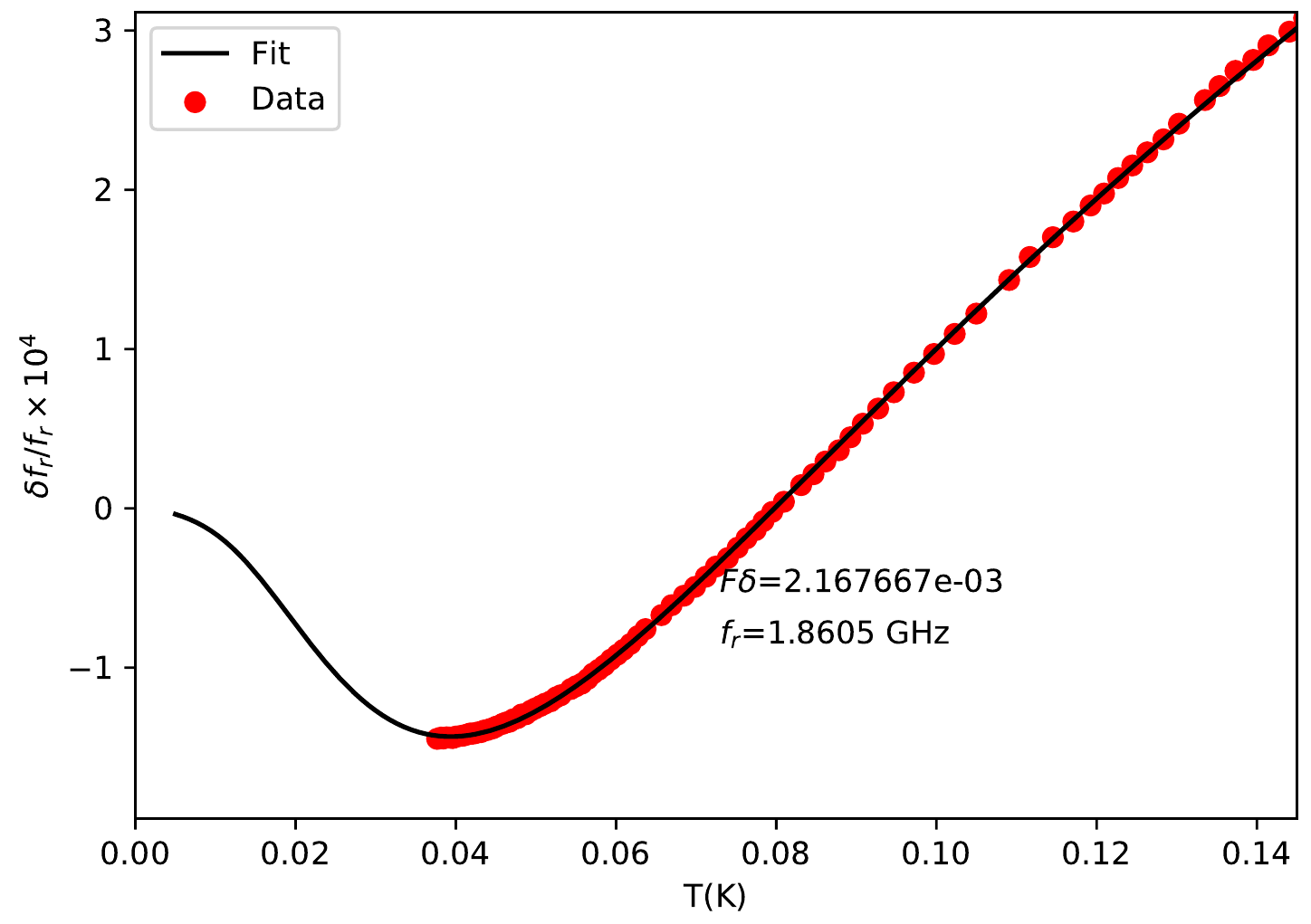}
\caption{Resonator frequency shift $f(T)$ vs. temperature $T$ and the fit to a TLS model. The fit $F\delta_{\mathrm{TLS}}^0$ is $2.2\times 10^{-3}$ for this resonator.}
\label{fig:fd_fit}
\end{figure}

\begin{figure}[!ht]
\centering
\includegraphics[width=\linewidth]{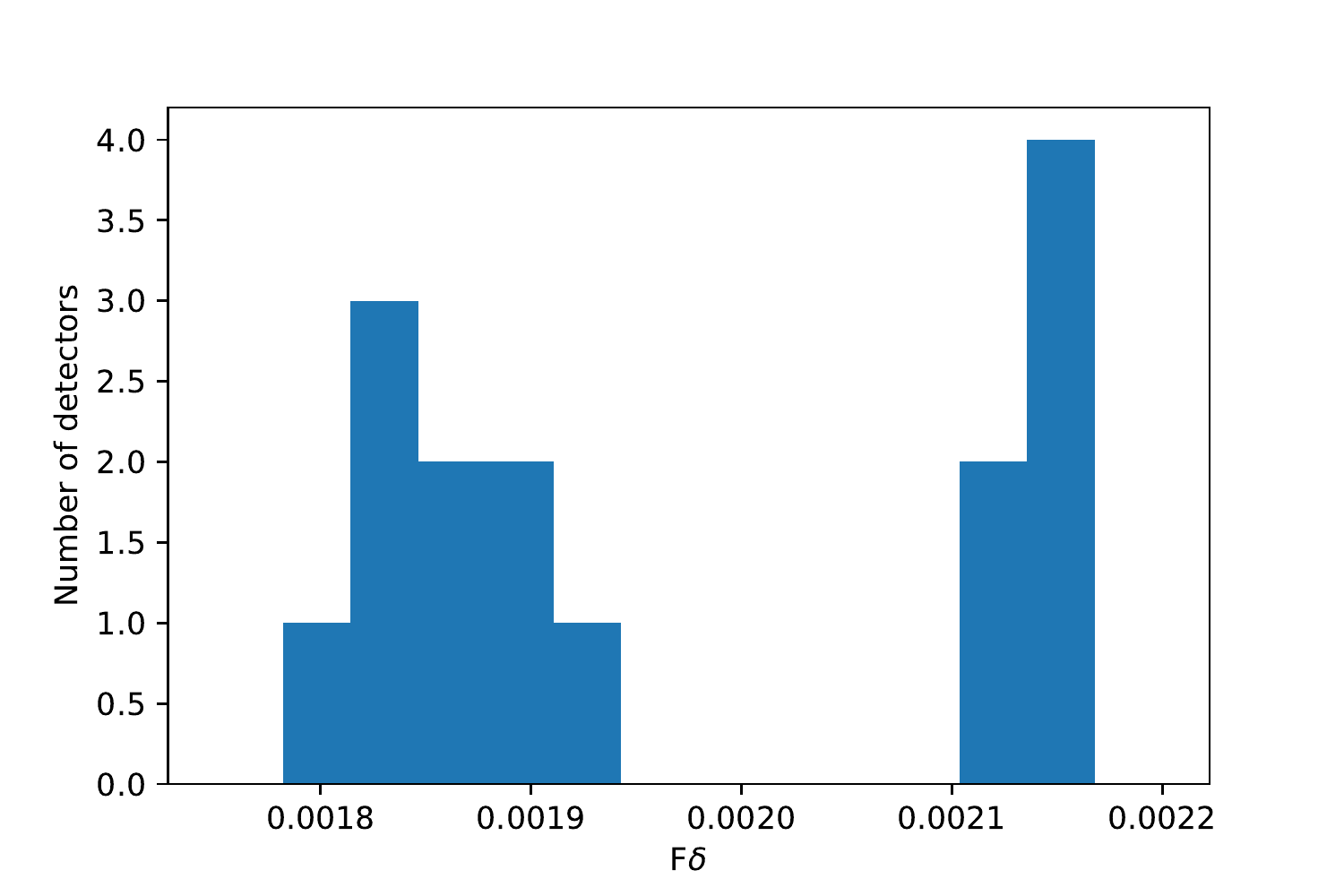}
\caption{Histogram showing the measured $F\delta_{\mathrm{TLS}}^0$ for all resonators fitting to the TLS model. The two groups belong to two different detector chips. Albeit the small difference, the two chips both show $F\delta_{\mathrm{TLS}}^0$  at the $2\times 10^{-3}$ level. The statistical error within each chip is within $10^{-4}$.  }
\label{fig:fd_distribution}
\end{figure}

We measured the resonance frequencies for our resonators using a standard vector network analyzer (VNA) setup with low-temperature attenuators and amplifiers and a room-temperature last-stage amplifier. A sample measurement was shown in Fig. \ref{fig:fd_fit} for a detector with SiN$_x$ dielectric. We fit for the model in Eq. (\ref{eq:tls_model}) to extract $F\delta_{\mathrm{TLS}}^0$. The fit result of $\delta_{\mathrm{TLS}}^0$ is $2.2\times 10^{-3}$ for this resonator, which is representative of detectors on the same chip. Fig. \ref{fig:fd_distribution} summarizes the distribution of $F\delta_{\mathrm{TLS}}^0$ from two chips, which corresponds to the two groups in the histogram. $F\delta_{\mathrm{TLS}}^0$ within each chip is uniform within $\pm 10^{-4}$ and the systematic shift between the two chips is also at the $10^{-4}$ level. To conclude, this measurement yields $\delta_{\mathrm{TLS}}^0=(2.0\pm 0.2)\times 10^{-3}$ for SiN$_x$.

\section{Dielectric Loss from $1/Q_{\mathrm{TLS}}$ at 1-2~GHz}
\label{sect:method3}
We can achieve quality factors at a few hundred thousand for resonators using interdigitated capacitors but only get quality factors at O($10^3$) level for resonators with parallel-plate capacitors and the same metal targets. This result suggests that most loss comes from TLS in SiN${_x}$ dielectric layer and that the internal quality factor $Q_i$ is limited by TLS loss. We can then use $1/Q_i$ to approximate dielectric loss. 
 
However, one caveat for this measurement is that the driving power can saturate the TLS states and lead to underestimated $F\delta_{\mathrm{TLS}}^0$. From \cite{gao2008physics}, the resonator loss $\delta_{\mathrm{TLS}}$ depends on power and temperature:
\begin{equation}
\delta_{\mathrm{TLS}} = F\delta_{\mathrm{TLS}}^0 \frac{\mathrm{tanh}(\frac{\hbar\omega}{2k_B T})}{\sqrt{1+\frac{\langle n \rangle}{n_c}}}, 
\end{equation}
where $\langle n \rangle$ is the average photon number in the resonator, and $n_c$ is the TLS saturation photon number. $\langle n \rangle$ is proportional to power, so $1/Q_i\sim 1/Q_{\mathrm{TLS}}= \delta_{\mathrm{TLS}}  \sim (1+P/P_c)^{-1/2} F\delta_{\mathrm{TLS}}^0$, where $P$ is the power and $P_c$ is the power for TLS saturation. We need the power sufficiently low to use $1/Q_i$ as a proxy for $F\delta_{\mathrm{TLS}}^0$. 

\begin{figure}[!ht]
\centering
\includegraphics[width=\linewidth]{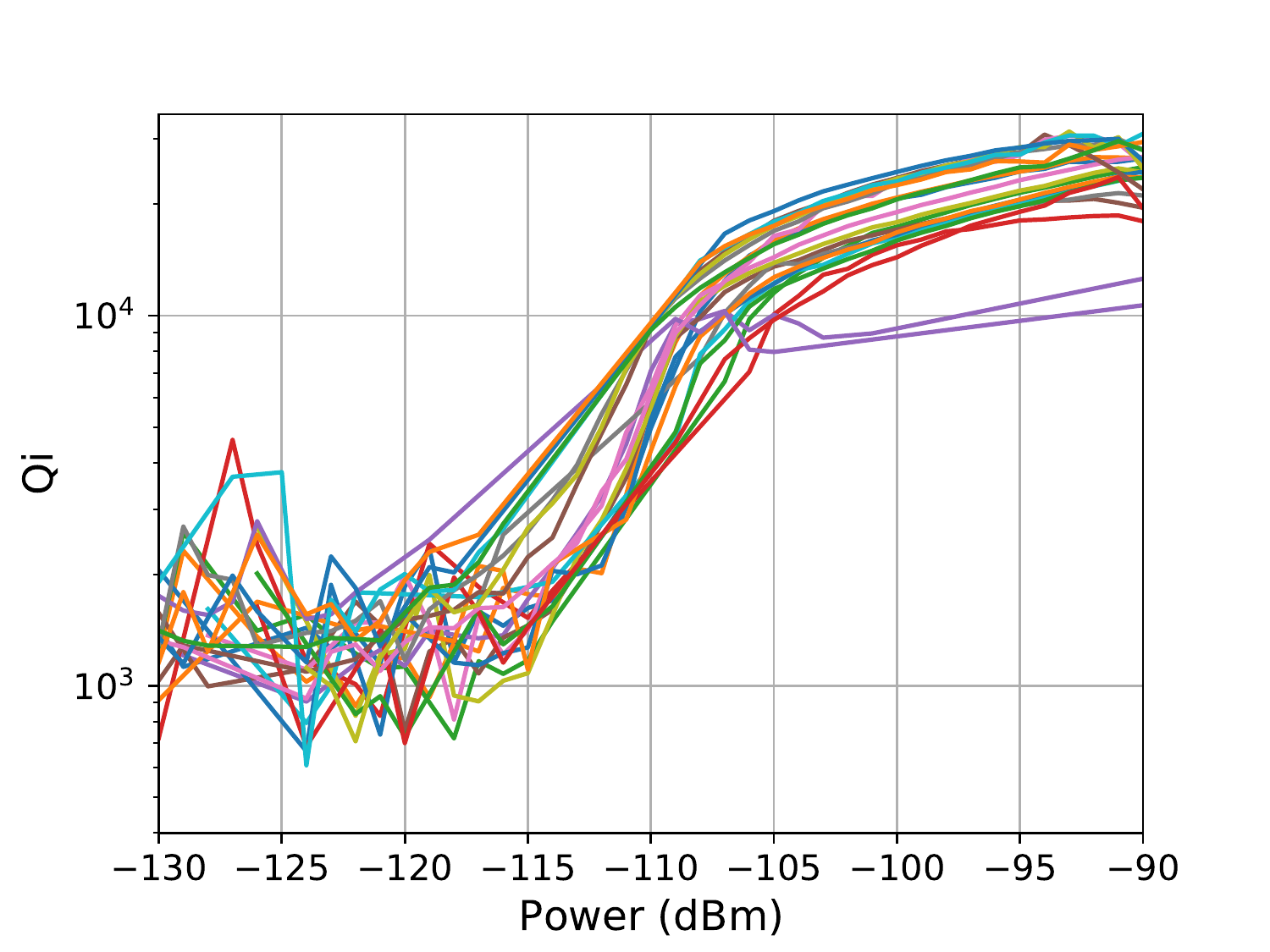}
\caption{ Resonator quality factor vs. driving power. Each curve is a different detector. As the power increases, more TLS states become saturated, leading to a higher $Q_i$. At powers below  $-$125~dBm, the VNA data becomes nosier, making it harder to constrain $Q_i$ in the low power limit. Above $-$105~dBm, the $Q_i$ is limited by other loss mechanisms, such as loss in the metal. }
\label{fig:qi_vs_pwr}
\end{figure}
\begin{figure}[!ht]
\centering
\includegraphics[width=\linewidth]{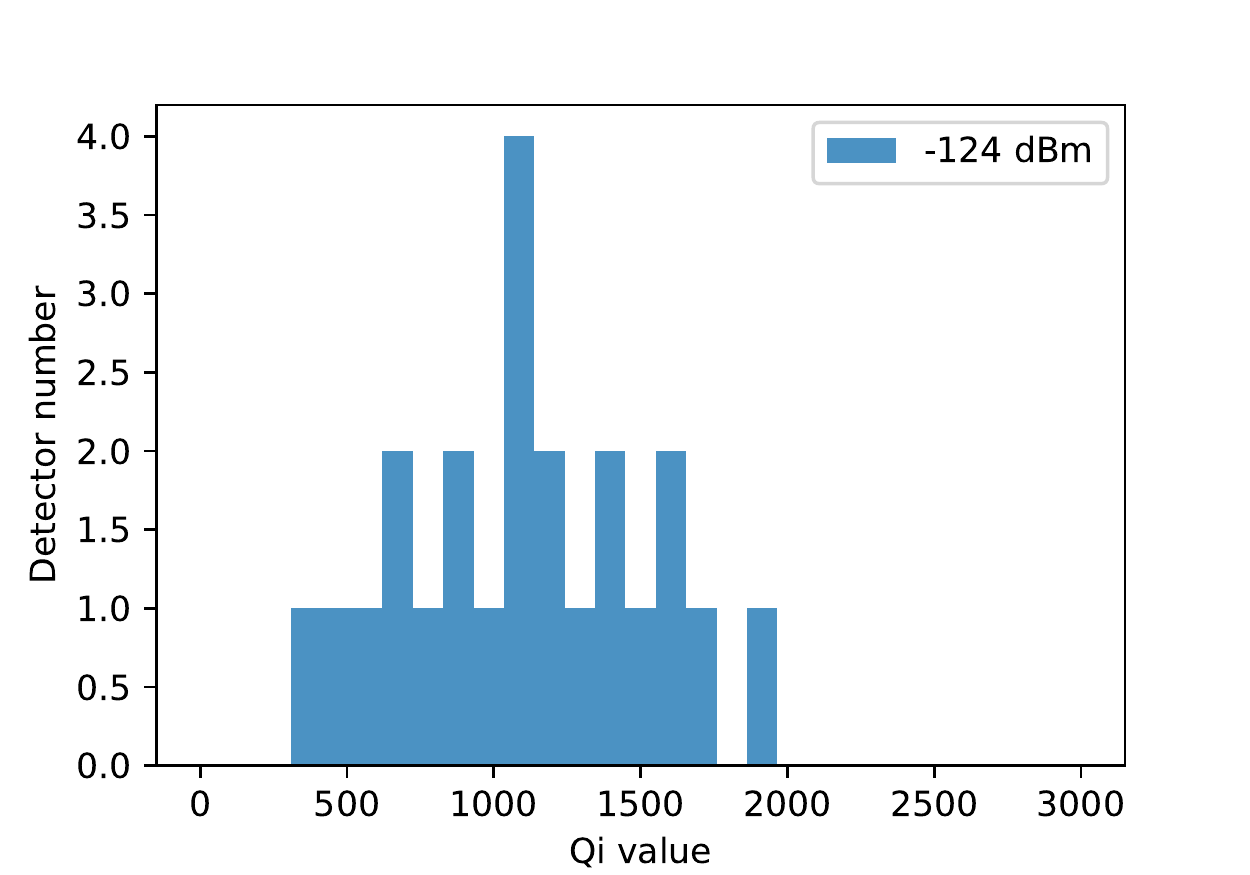}
\caption{$Q_i$ distribution at $-$124~dBm feedline power. }
\label{fig:qi_distribution}
\end{figure}

Fig. \ref{fig:qi_vs_pwr} shows the resonator $Q_i$ vs. power at the feedline for all detectors we measured. We observe an increasing trend of $Q_i$ vs. power due to TLS saturation. At low powers below $-125$~dBm, the VNA loses S/N. As a result, we cannot probe into the low-power limit and can only obtain a higher limit for $Q_i$ (without TLS saturation) and a lower limit for $F\delta_{\mathrm{TLS}}^0$. The histogram of resonator $Q_i$ at $-124$~dBm is in Fig. \ref{fig:qi_distribution} and centers around {$Q_i \sim 1000 \pm 100$}, corresponding to a constraint of $\delta_{\mathrm{TLS}}^{0 }\mathrm{\gtrsim 1.0\times 10^{-3}}$. {The results are consistent with those from fitting to the TLS model (Section \ref{sect: tls_model})} in the same order of magnitude. We tried fitting the $Q_i$ dependence on power to the $(1+P/P_c)^{-1/2}$ model, but there is strong degeneracy between saturation power $P_c$ and $F\delta_{\mathrm{TLS}}^0$. To improve this method, we can switch to a lower-noise amplifier or increase the integration time during data collection. {We also note that the high-power microwave $Q_i$ of our microstrip with SiN$_x$ is lower than that reported by other groups \cite{endo2013chip, paik2010reducing}. The $Q_i$ for resonators with aluminum inductors used in this paper still has a positive slope vs. power at $-$90dBm, indicating their $Q_i$ is still being limited by dielectric loss there. However, these resonators started to bifurcate at $\sim -$90dBm. For all-Nb resonators with higher bifurcation power, we noticed that the $Q_i$ can reach $10^5$ at $-$70~dBm. For this paper, we are only interested in the low-power limit of the $Q_i$ set by the dielectric loss. The other factors that limit our high-power $Q_i$ are orders of magnitude lower than the TLS loss. }

\section{Differential Optical Efficiency at 150~GHz}

\begin{figure}[!ht]
\centering
\includegraphics[width=\linewidth]{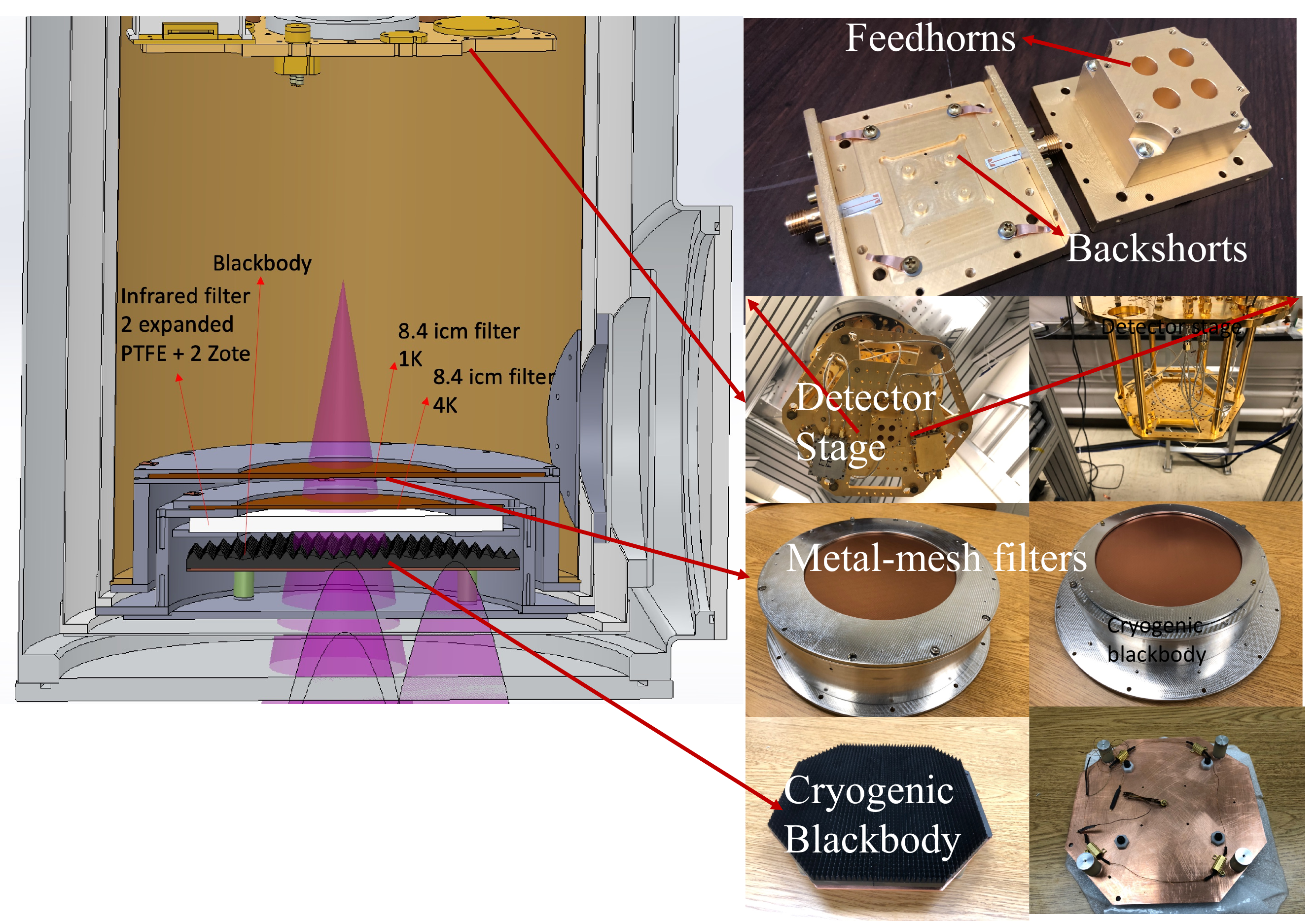}
\caption{Optical testing setup. The cryogenic blackbody is mounted on a copper plate with four heaters and thermometers for temperature control. An infrared filter attenuates the power above 1~THz, and two metal-mesh filters at 4~K and 1~K define a low-pass frequency window below 280~GHz. The detectors are mounted in a sample box with feedhorns and backshorts. We installed the optical components in a Bluefors dilution fridge shown on the left. The base temperature of the fridge is 8~mK. The optical window on the right of the CAD model came with the fridge and was not used for this paper. }
\label{fig:coupling_setup}
\end{figure}

The third method compares optical efficiency for detectors coupled through different lengths of microstrip lines, which requests an mm-wave optical coupling setup. A temperature-controlled cryogenic blackbody provides a method to sweep the incident power level, and is coupled to the detector sample box through low-pass metal-mesh filters and infrared blockers (Fig. \ref{fig:coupling_setup}). 

\begin{figure}[!ht]
\centering
\includegraphics[width=0.8\linewidth]{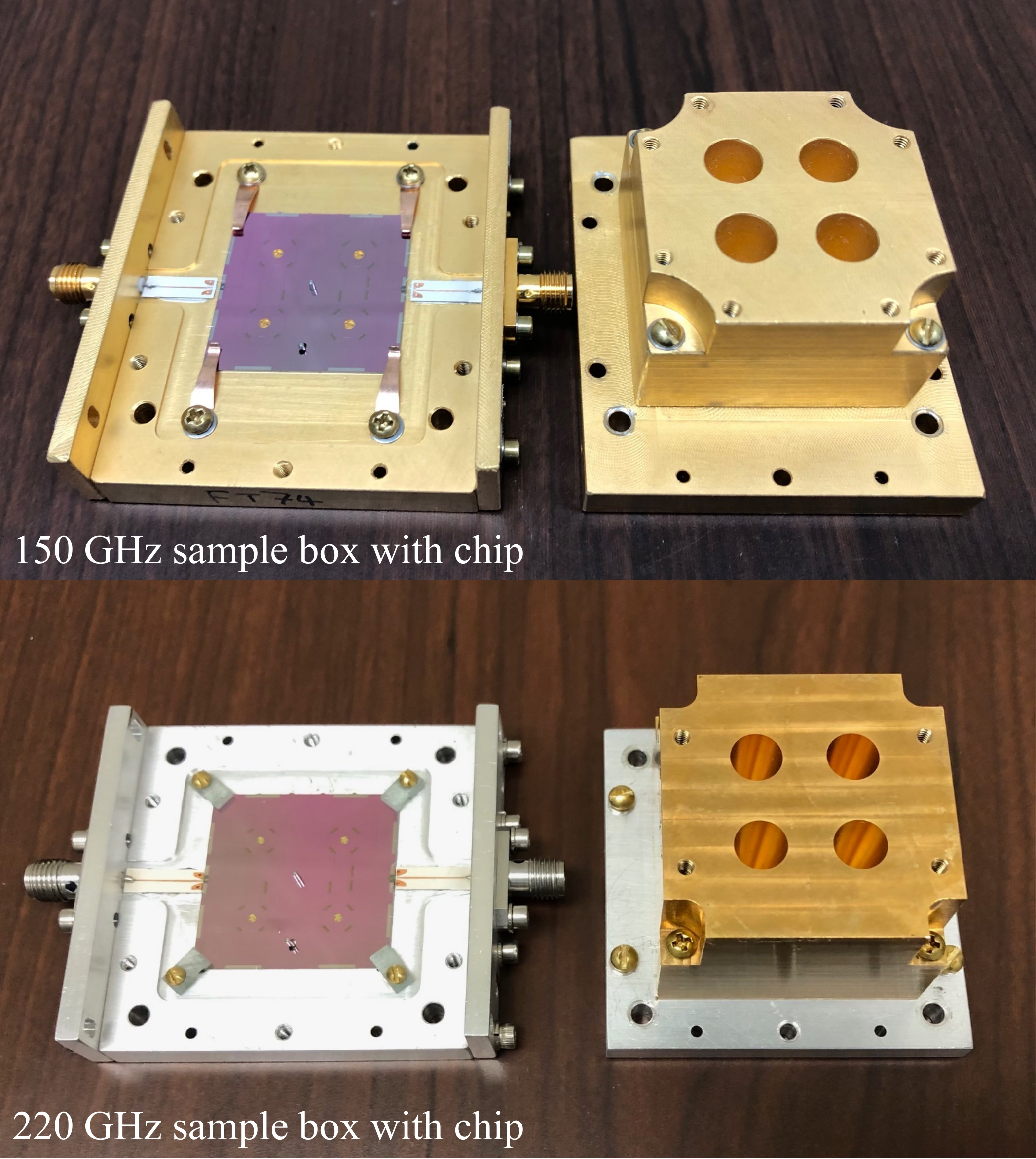}
\caption{150 and 220~GHz sample boxes with detector chips installed. Left of each subfigure is the box bottom with the detector chip. Right is the box lid with feedhorns.   }
\label{fig:sample_boxes}
\end{figure}

The optical setup we built is compatible with both 150 and 220~GHz {optical detector chips}. Fig. \ref{fig:sample_boxes} shows the detectors and their boxes. The 150 and 220~GHz chips were fabricated and diced from the same silicon wafer. 

\begin{figure}[!ht]
\centering
\includegraphics[width=\linewidth]{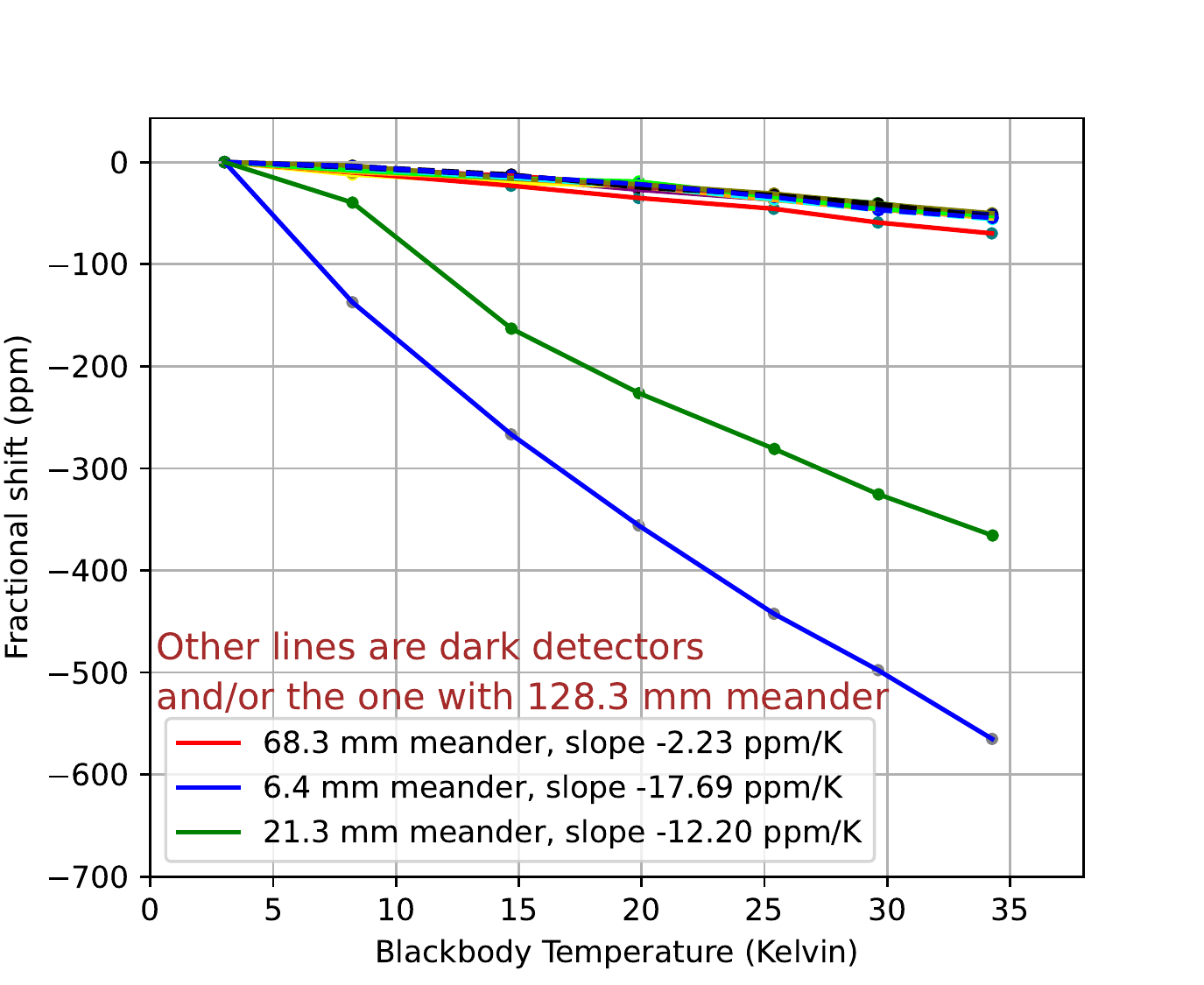}
\caption{Frequency shift vs. cold load temperature. We have four optical detectors coupled to 6.4, 21.3, 68.3, and 128.3mm meanders. The detector with the longest length (128.3mm) did not show a clear optical response due to too much attenuation. Dark detectors showed some response to cold load temperature due to chip heating.}
\label{fig:optical_response}
\end{figure}

To measure the optical response of the detectors, we heat the blackbody to different temperatures $T_{BB}$ and record the resonator frequency shift $d f/f_0$ vs. $T_{BB}$. Here $df$ is the frequency shift, and $f_0$ is the resonant frequency. We took measurements for the 150~GHz device and show the results in Fig. \ref{fig:optical_response}, where {we identified three optical detectors with decreasing optical response corresponding to microstrip meanders lengths of 6.4, 21.3, and 68.3~mm}. The detector with the longest microstrip length of 128.3~mm has low optical response comparable to other dark detectors, suggesting its optical power is too weak from too much attenuation. The frequencies for the dark detectors also shifted slightly due to the chip heating. We keep the detector stage at a constant $T_{\mathrm{stage}}=$247~mK where $df/dT_{\mathrm{stage}}\sim0$ to minimize the detector response to base temperature variation. However, as the blackbody radiates more power to the device, the temperature gradient between the chip and the stage PID thermometer can increase as a function of blackbody temperature, causing detector frequency shifts. We successfully reduced this effect by fine-tuning the stage PID temperature such that $df/dT_{\mathrm{stage}}$ is closer to zero. But since the device temperature ($T_{\mathrm{chip}}$) is not a constant and depends on the blackbody temperature $T_{BB}$, $df/dT_{\mathrm{stage}}$ cannot always stay zero at varying $T_{BB}$ even if we PID control $T_{\mathrm{stage}}$ to a constant. We will always have a small amount of $T_{\mathrm{chip}}$ response, which can be monitored and corrected for using dark detectors. 

The next step is to extract the dielectric loss from the measured optical responses. We use 
\begin{equation}
\label{eq:delta_efficiency}
\tan \delta=\frac{c}{2 \pi L \sqrt{\xi_r} f_{\text{opt}}} \ln \left(\frac{\eta_{\text {long }}}{\eta_{\text {short }}}\right)
\end{equation}
 to calculate the loss through unit microstrip line length \cite{hubmayr2022optical}. Here $c$ is the speed of light, $L$ is the length difference, $f_{\text{opt}}$ is the optical band central frequency (150 or 220~GHz depending on the chip style), {$\eta_{\text {short}}$($\eta_{\text{long }}$)} is the optical efficiency for the short (long) meander, and {permittivity} $\xi_r=7.0$. The fractional frequency shift can be decomposed into 
 \begin{equation}
d f/f_0=P_{\mathrm{o p t}} \frac{d f/f_0}{d n_{\mathrm{q p}}} \frac{d n_{\mathrm{q p}}}{P_{\mathrm{o p t}}}=P_{\mathrm{o p t}} \frac{d f/f_0}{d n_{\mathrm{q p}}} \frac{\eta \tau_{\mathrm{q p}}}{\Delta V_L},
 \end{equation}
where $P_{opt}$ is the optical power, $n_{\mathrm{qp}}$ is the quasiparticle density, $\eta$ is the optical efficiency including line loss and pair-breaking efficiency, $\tau_{qp}$ is the quasiparticle lifetime, $\Delta$ is the bandgap, and $V_L$ is the inductor volume \cite{baselmans2008noise}. We designed our inductor volume $V_L$ to be the same for all detectors. $\tau_{qp}$ and $\Delta$ are the same for the same wafer. $P_{\mathrm{opt}}$ should be the same for the same feedhorn, antenna, and {blackbody illumination that covers the entire beam of the detectors}. We measured $df/(f_0dn_{\mathrm{qp}})$ from resonant frequency vs. operating temperature. $n_{qp}$ depends on the operating temperature $n_{\mathrm{q p}}=2 N_0 \sqrt{2 \pi k_B T \Delta} \exp \left(-\Delta / k_B T\right)$ \cite{de2011number}, where $N_0$ is the single spin density of states at the Fermi level. The difference for $df/(f_0dn_{\mathrm{qp}})$  measured within 0.7\% for different detectors.  Since all other terms are approximately the same, the $df/f_0$ vs. optical power (scaled with blackbody temperature $T_{BB}$) slopes can be directly divided to get the optical efficiency ratio, or $\eta_{\text {long }}/\eta_{\text {short }}$ in Eq. \ref{eq:delta_efficiency}. We get $\tan\delta\sim$0.003 from 21.3~mm/ 6.4~mm detector responses and 0.005 from 68.3~mm/ 6.4~mm. On average $\tan\delta\sim (4\pm 1)\times 10^{-3}$. {We note that our loss tangent is larger than reported by \cite{hailey2014optical}. Our group is currently developing SiN$_x$ dielectrics using several deposition methods: chemical vapor deposition (CVD), ion beam-assisted sputtering (IBAS), and PECVD. We have measured microwave loss tangents as low as $2x10^{-4}$ \cite{margarita2023} in silicon-rich SiN$_x$ films deposited by PECVD. A future paper will present additional details on the dielectric optimization for mm-wave loss tangent; the focus of this paper is on our measurement methodology. }.

\section{Measurement systematics}
The systematics for our optical measurements include the chip heating effect discussed in Section \ref{sect:method3}, optical leakage, and optical coupling variation among different pixels. We can control chip heating by choosing an appropriate stage PID temperature where the $df/dT_{\mathrm{stage}}$ responsivity is low and using the dark detectors to estimate a correction. One concern about our optical coupling is that different antennas will have slightly different coupling efficiency if they are misaligned under the feedhorns. A series of simulations suggest that the optical efficiency variation should be within ten percent if we can align the antennas and feedhorns within 0.1~mm. Optical leakage mitigation is a more challenging task  \cite{gualtieri2023}. Below we discuss the leakage measurements and a method for reducing the optical leakage with absorber chips.

\begin{figure}[!ht]
\centering
\includegraphics[width=\linewidth]{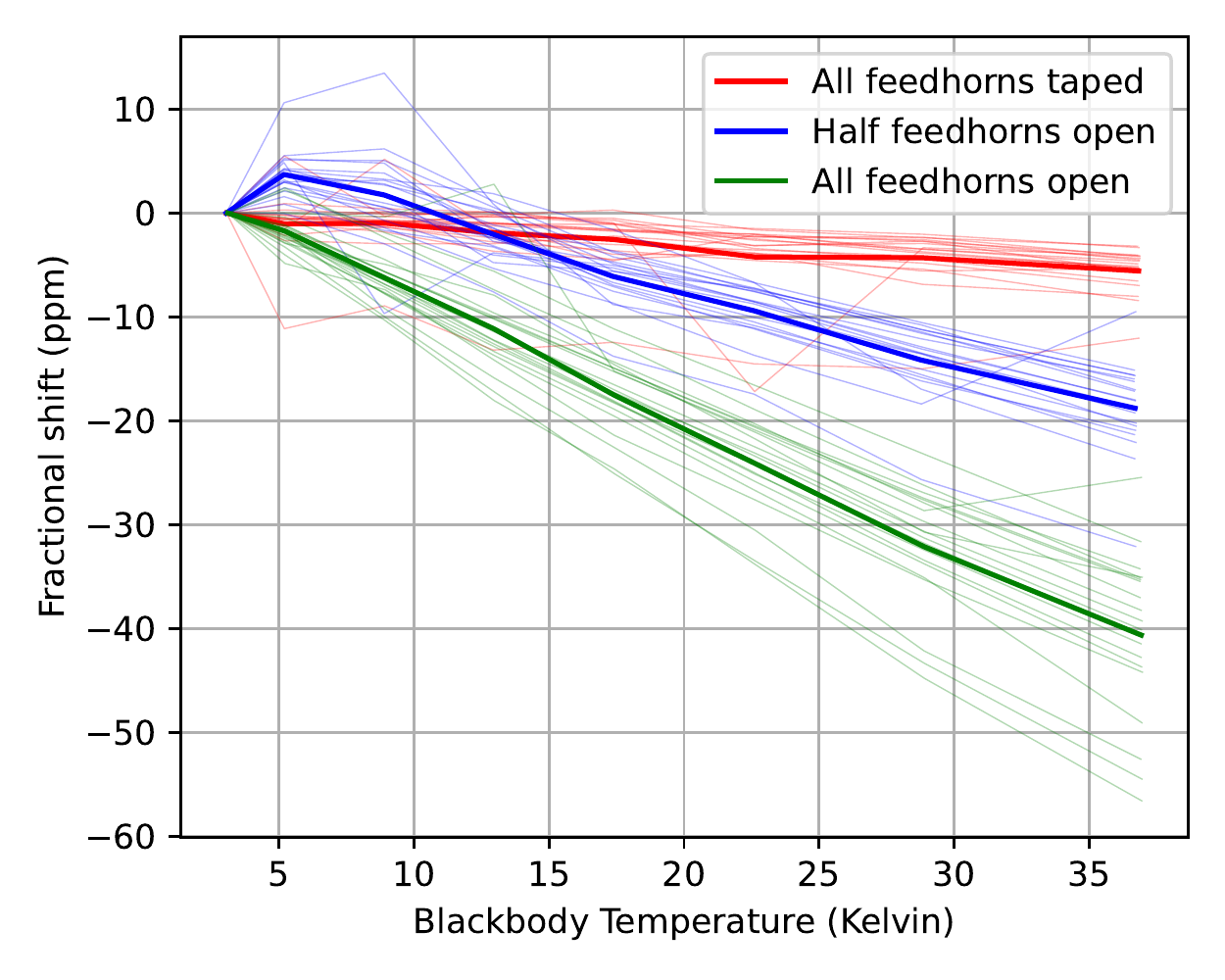}
\caption{Fractional frequency shift vs. blackbody temperature for dark detectors with different numbers of feedhorns taped. The response levels for most dark detectors are proportional to the number of open feedhorns. {The thick line is the average over all detectors (thinner lines).}
}
\label{fig:dark_pickup}
\end{figure}

\begin{figure}[!ht]
\centering
\includegraphics[width=\linewidth]{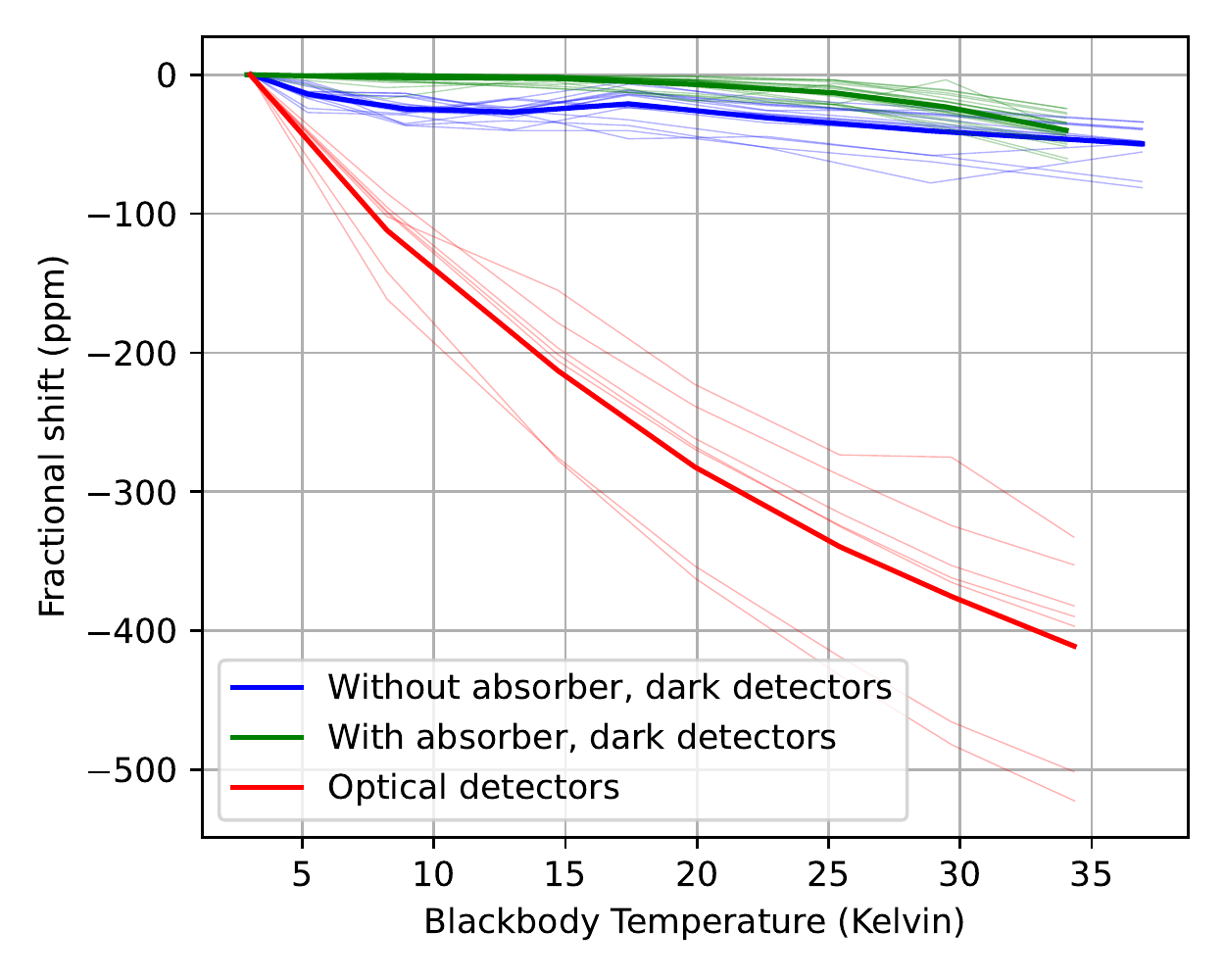}
\caption{Fractional frequency shift vs. blackbody temperature for the dark detectors without absorber chips, dark detectors with absorber chips, and optical detectors. The dark pickup is at $\sim$10\% {of the response level for an optical detector} and can be mitigated by installing absorber chips. The thick line is the average over all detectors (thinner lines).    }
\label{fig:absorber_suppression}
\end{figure}

We observed dark detector response at 10\% compared to optical detectors before any leakage mitigation. Fig. \ref{fig:dark_pickup} shows the response level for the dark detectors. We measured the optical response for detectors on the same device with different numbers of feedhorns covered over the metallic tape. The responses of all dark detectors were $\sim$1~ppm/K with all feedhorns open and uniformly reduced in half after we covered two of the four feedhorns. The amount of reduction is independent of the detector location on the device. When we covered all feedhorns, the detector responses were all close to zero, with a small slope likely caused by chip heating. The measurements are consistent with a model where optical leakage through the bottom of the feedhorn randomly scatter within the sample box and can terminate at any inductors. The leaked photons can break Cooper pairs in the inductor and shift the resonant frequency. The first data point of the blue line in Fig.~\ref{fig:dark_pickup} was an outlier caused by stage temperature variation when we started the temperature PID control. 

\begin{figure}[!ht]
\centering
\includegraphics[width=0.6\linewidth]{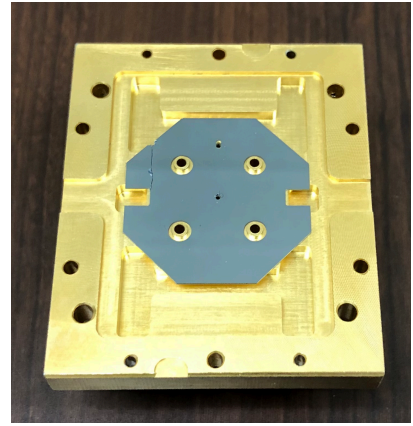}
\caption{An absorber chip is mounted in the box lid for terminating spurious photons. An additional absorber chip is underneath the detector chip and is not shown in this picture. The side with absorbing patterns should face the detector chip for both absorber chips. The other side of the absorber chip has a metal layer that reflects the radiation to enhance the absorption. }
\label{fig:absorber_chip}
\end{figure}

To address this parasitic optical leakage, we designed and fabricated  dedicated absorbing chips and placed them both above and below the detector chip to absorb spurious photons. Each absorber chip comprises a piece of silicon with a tuned absorbing structure that is patterned into an aluminium thin film. Fig.~\ref{fig:absorber_suppression} shows the responses for optical detectors and dark detectors with and without the absorber chips. The dark response was about 1~ppm/K before adding the absorber chips and was reduced to below measurement noise level afterward. The nonlinear shift for the green curve at $T_{BB}>25$~K is likely caused by chip heating. A picture of the top absorber device is in Fig.~\ref{fig:absorber_chip}.

\section{Conclusions}

This paper presents dielectric loss measurements for SiN$_x$ using three different methods covering millimeter to centimeter wavelengths with the same test structure. At $\sim$1-2~GHz, we measured SiN$_x$ loss to be $2\times10^{-3}$ using the TLS fitting method and  $\mathrm{\gtrsim 1.0\times 10^{-3}}$ by inverting $Q_i$ at low excitation power. At 150~GHz, we measured the loss to be $\sim4\times10^{-3}$ using the differential optical efficiency method. The loss at 150~GHz is a factor of $\sim$2 larger compared to the loss at $\sim$1-2 GHz, which is consistent with NIST measurements \cite{hubmayr2022optical}. Next, we will measure dielectric loss for a few more materials, such as SiO$_x$, silicon-rich SiN$_x$, and amorphous silicon. Early results showed a promising loss tangent of $2\times10^{-4}$ at ~1~GHz for silicon-rich SiN$_x$. We still need to understand why the loss tangent at 150~GHz is larger than at $\sim$1 GHz. One possibility is other loss mechanisms beyond TLS at mm-wave, which can be tested using samples with lower loss, such as silicon-rich SiN$_x$. We have made 220~GHz samples and boxes and plan to extend the measurements to 220~GHz. 

\section{Acknowledgements}
Work at Argonne, including the use of the Center for Nanoscale Materials, an Office of Science user facility, was supported by the U.S. Department of Energy, Office of Science, Office of Basic Energy Sciences, and Office of High Energy Physics, under Contract No. DE-AC02-06CH11357. Zhaodi Pan is supported by ANL under the award LDRD-2021-0186.
\bibliographystyle{IEEEtran}
\bibliography{paper_reference.bib}

\begin{thebibliography}{10}
\providecommand{\url}[1]{#1}
\csname url@samestyle\endcsname
\providecommand{\newblock}{\relax}
\providecommand{\bibinfo}[2]{#2}
\providecommand{\BIBentrySTDinterwordspacing}{\spaceskip=0pt\relax}
\providecommand{\BIBentryALTinterwordstretchfactor}{4}
\providecommand{\BIBentryALTinterwordspacing}{\spaceskip=\fontdimen2\font plus
\BIBentryALTinterwordstretchfactor\fontdimen3\font minus
  \fontdimen4\font\relax}
\providecommand{\BIBforeignlanguage}[2]{{%
\expandafter\ifx\csname l@#1\endcsname\relax
\typeout{** WARNING: IEEEtran.bst: No hyphenation pattern has been}%
\typeout{** loaded for the language `#1'. Using the pattern for}%
\typeout{** the default language instead.}%
\else
\language=\csname l@#1\endcsname
\fi
#2}}
\providecommand{\BIBdecl}{\relax}
\BIBdecl

\bibitem{abitbol2017cmb}
M.~H. Abitbol, Z.~Ahmed, D.~Barron, R.~B. Thakur, A.~N. Bender, B.~A. Benson,
  C.~A. Bischoff, S.~A. Bryan, J.~E. Carlstrom, C.~L. Chang \emph{et~al.},
  ``Cmb-s4 technology book,'' \emph{arXiv preprint arXiv:1706.02464}, 2017.

\bibitem{hailey2016low}
S.~Hailey-Dunsheath, E.~Shirokoff, P.~Barry, C.~Bradford, S.~Chapman, G.~Che,
  J.~Glenn, M.~Hollister, A.~Kov{\'a}cs, H.~LeDuc \emph{et~al.}, ``Low noise
  titanium nitride kids for superspec: A millimeter-wave on-chip
  spectrometer,'' \emph{Journal of Low Temperature Physics}, vol. 184, pp.
  180--187, 2016.

\bibitem{martinis2005decoherence}
J.~M. Martinis, K.~B. Cooper, R.~McDermott, M.~Steffen, M.~Ansmann, K.~Osborn,
  K.~Cicak, S.~Oh, D.~P. Pappas, R.~W. Simmonds \emph{et~al.}, ``Decoherence in
  josephson qubits from dielectric loss,'' \emph{Physical review letters},
  vol.~95, no.~21, p. 210503, 2005.

\bibitem{li2013improvements}
D.~Li, J.~Gao, J.~Austermann, J.~Beall, D.~Becker, H.-M. Cho, A.~E. Fox,
  N.~Halverson, J.~Henning, G.~Hilton \emph{et~al.}, ``Improvements in silicon
  oxide dielectric loss for superconducting microwave detector circuits,''
  \emph{IEEE transactions on applied superconductivity}, vol.~23, no.~3, pp.
  1\,501\,204--1\,501\,204, 2013.

\bibitem{ye2019low}
Z.~Ye, A.~F{\"u}l{\"o}p, {\'O}.~B. Helgason, P.~A. Andrekson \emph{et~al.},
  ``Low-loss high-q silicon-rich silicon nitride microresonators for kerr
  nonlinear optics,'' \emph{Optics Letters}, vol.~44, no.~13, pp. 3326--3329,
  2019.

\bibitem{defrance2022low}
F.~Defrance, S.~Shu, A.~Beyer, J.~Wan, J.~Sayers, and S.~Golwala, ``Low
  intrinsic tls loss hydrogenated amorphous silicon,'' in \emph{Millimeter,
  Submillimeter, and Far-Infrared Detectors and Instrumentation for Astronomy
  XI}.\hskip 1em plus 0.5em minus 0.4em\relax SPIE, 2022, p. PC121900D.

\bibitem{buijtendorp2021hydrogenated}
B.~Buijtendorp, S.~Vollebregt, K.~Karatsu, D.~Thoen, V.~Murugesan,
  K.~Kouwenhoven, S.~H{\"a}hnle, J.~Baselmans, and A.~Endo, ``Hydrogenated
  amorphous silicon carbide: a low-loss deposited dielectric for microwave to
  submillimeter wave superconducting circuits,'' \emph{arXiv preprint
  arXiv:2110.03500}, 2021.

\bibitem{hubmayr2022optical}
J.~Hubmayr, P.~Ade, A.~Adler, E.~Allys, D.~Alonso, K.~Arnold, D.~Auguste,
  J.~Aumont, R.~Aurlien, J.~Austermann \emph{et~al.}, ``Optical
  characterization of omt-coupled tes bolometers for litebird,'' \emph{Journal
  of Low Temperature Physics}, pp. 1--13, 2022.

\bibitem{buijtendorp2022hydrogenated}
B.~Buijtendorp, S.~Vollebregt, K.~Karatsu, D.~Thoen, V.~Murugesan,
  K.~Kouwenhoven, S.~H{\"a}hnle, J.~Baselmans, and A.~Endo, ``Hydrogenated
  amorphous silicon carbide: A low-loss deposited dielectric for microwave to
  submillimeter-wave superconducting circuits,'' \emph{Physical Review
  Applied}, vol.~18, no.~6, p. 064003, 2022.

\bibitem{hahnle2021superconducting}
S.~H{\"a}hnle, K.~Kouwenhoven, B.~Buijtendorp, A.~Endo, K.~Karatsu, D.~Thoen,
  V.~Murugesan, and J.~Baselmans, ``Superconducting microstrip losses at
  microwave and submillimeter wavelengths,'' \emph{Physical Review Applied},
  vol.~16, no.~1, p. 014019, 2021.

\bibitem{hood2022testing}
J.~Hood, P.~Barry, T.~Cecil, C.~Chang, J.~Li, S.~Meyer, Z.~Pan, E.~Shirokoff,
  and A.~Tang, ``Testing low-loss microstrip materials with mkids for microwave
  applications,'' \emph{Journal of Low Temperature Physics}, pp. 1--7, 2022.

\bibitem{anderson1972anomalous}
P.~W. Anderson, B.~I. Halperin, and C.~M. Varma, ``Anomalous low-temperature
  thermal properties of glasses and spin glasses,'' \emph{Philosophical
  Magazine}, vol.~25, no.~1, pp. 1--9, 1972.

\bibitem{phillips1987two}
W.~A. Phillips, ``Two-level states in glasses,'' \emph{Reports on Progress in
  Physics}, vol.~50, no.~12, p. 1657, 1987.

\bibitem{gao2008physics}
J.~Gao, \emph{The physics of superconducting microwave resonators}.\hskip 1em
  plus 0.5em minus 0.4em\relax California Institute of Technology, 2008.

\bibitem{gao2008equivalence}
J.~Gao, J.~Zmuidzinas, A.~Vayonakis, P.~Day, B.~Mazin, and H.~Leduc,
  ``Equivalence of the effects on the complex conductivity of superconductor
  due to temperature change and external pair breaking,'' \emph{Journal of Low
  Temperature Physics}, vol. 151, no.~1, pp. 557--563, 2008.

\bibitem{endo2013chip}
A.~Endo, C.~Sfiligoj, S.~Yates, J.~Baselmans, D.~Thoen, S.~M.~H. Javadzadeh,
  P.~Van~der Werf, A.~Baryshev, and T.~Klapwijk, ``On-chip filter bank
  spectroscopy at 600--700 ghz using nbtin superconducting resonators,''
  \emph{Applied Physics Letters}, vol. 103, no.~3, p. 032601, 2013.

\bibitem{paik2010reducing}
H.~Paik and K.~D. Osborn, ``Reducing quantum-regime dielectric loss of silicon
  nitride for superconducting quantum circuits,'' \emph{Applied Physics
  Letters}, vol.~96, no.~7, p. 072505, 2010.

\bibitem{baselmans2008noise}
J.~Baselmans, S.~Yates, R.~Barends, Y.~Lankwarden, J.~Gao, H.~Hoevers, and
  T.~Klapwijk, ``Noise and sensitivity of aluminum kinetic inductance detectors
  for sub-mm astronomy,'' \emph{Journal of Low Temperature Physics}, vol. 151,
  no. 1-2, pp. 524--529, 2008.

\bibitem{de2011number}
P.~De~Visser, J.~Baselmans, P.~Diener, S.~Yates, A.~Endo, and T.~Klapwijk,
  ``Number fluctuations of sparse quasiparticles in a superconductor,''
  \emph{Physical review letters}, vol. 106, no.~16, p. 167004, 2011.

\bibitem{hailey2014optical}
S.~Hailey-Dunsheath, P.~Barry, C.~Bradford, G.~Chattopadhyay, P.~Day, S.~Doyle,
  M.~Hollister, A.~Kovacs, H.~LeDuc, N.~Llombart \emph{et~al.}, ``Optical
  measurements of superspec: A millimeter-wave on-chip spectrometer,''
  \emph{Journal of Low Temperature Physics}, vol. 176, pp. 841--847, 2014.

\bibitem{margarita2023}
M.~Lisovenko, Z.~Pan, P.~S. Barry, T.~Cecil, C.~L. Chang, K.~R. Dibert,
  R.~Gualtieri, J.~Li, V.~Novosad, G.~Wang, and V.~Yefremenko, ``Low-loss
  dielectrics for high frequency components of superconducting detectors,''
  \emph{IEEE Transactions of Superconductivity (this special issue)}, p. N. A.,
  2023.

\bibitem{gualtieri2023}
R.~Gualtieri, P.~Barry, T.~Cecil, A.~Bender, C.~Chang, J.~Hood, J.~Li, and
  M.~Lisovenko, ``Optical leakage mitigation in superconducting ortho-mode
  transducer arrays,'' \emph{Presented at ASC 2022, Honolulu, Hawaii, USA,
  October 2022, Paper number ASC2022-4EOr1A-05}, p. N. A., 2022.

\end{thebibliography}
\end{document}